\documentclass[twocolumn,nofootinbib,amsfonts,prd,aps]{revtex4}
\usepackage{graphicx}
\usepackage{subfig}
\usepackage{amsmath,amssymb,amsfonts,amsthm,mathrsfs}

\begin{document}

\title{Quantum self-gravitating collapsing matter in a quantum geometry}

\author{Miguel Campiglia$^{1}$, Rodolfo Gambini$^{1}$, Javier
  Olmedo$^{2}$, 
Jorge Pullin$^{2}$}
\affiliation {
1. Instituto de F\'{\i}sica, Facultad de Ciencias, 
Igu\'a 4225, esq. Mataojo, 11400 Montevideo, Uruguay. \\
2. Department of Physics and Astronomy, Louisiana State University,
Baton Rouge, LA 70803-4001}

\begin{abstract}
  The problem of how space-time responds to gravitating quantum matter
  in full quantum gravity has been one of the main questions that any
  program of quantization of gravity should address.  Here we analyze
  this issue by considering the quantization of a collapsing null
  shell coupled to spherically symmetric loop quantum gravity. We show that the
  constraint algebra of canonical gravity is Abelian both classically
  and when quantized using loop quantum gravity techniques. The
  Hamiltonian constraint is well defined and suitable Dirac
  observables characterizing the problem were identified at the
  quantum level. We can write the metric as a parameterized
  Dirac observable at the quantum level and study the physics of the
  collapsing shell and black hole formation. We show how the
  singularity inside the black hole is eliminated by loop quantum
  gravity and how the shell can traverse it. The construction is
  compatible with a scenario in which the shell tunnels into a baby
  universe inside the black hole or one in which it could emerge through a
  white hole.\end{abstract}
\maketitle

There has been recent progress in loop quantum gravity with spherical
symmetry. The key ingredient is the realization that through suitable
combinations and rescalings the constraint algebra can be made a Lie
algebra, both at a classical level and when quantized using loop
quantum gravity techniques \cite{spherical}. In vacuum the theory can
be solved in closed form including finding the space of physical
states annihilated by the constraints and suitable self-adjoint Dirac
observables characterizing the physics. The singularity that is
present inside black holes in the classical theory is replaced in the
quantum theory with a region with large curvatures, large fluctuations
of the curvature and high sensitivity to Planck scale microstructure,
so it cannot be well approximated by a semiclassical geometry. The
theory is well defined there and eventually one reaches a region
inside the black hole where space-time is again semiclassical. This
analysis has also been generalized to the electrovac case
\cite{electrovac}. Test (non-backreacting) shells were also studied in
this framework \cite{testshell}. In this letter we would like to
analyze the case of a self-gravitating shell. This is significant
since it is the first full quantum treatment of a midisuperspace model
including quantum matter in loop quantum gravity. It would also
provide a model for the emergence of a shell from a white hole, as
conjectured in several heuristic scenarios. It could also provide the
starting point for studying the backreaction of Hawking radiation on a
black hole, which cannot be properly studied with the previously
existing solutions, which were time independent \cite{mg}.


In a previous paper \cite{testshell} we studied the motion of
quantized null shells in a fixed quantum geometry background. In this
work we would like to consider the full backreaction of the shells on
the quantum geometry. We start from the Hamiltonian considered in
our previous manuscript. The context is spherically symmetric gravity
written in terms of Ashtekar variables. There are two pairs of
canonical variables $E^\varphi$, ${K}_\varphi$ and $E^x$, $K_x$, that
are related to the traditional canonical variables in spherical
symmetry $ds^2=\Lambda^2 dx^2+R^2 d\Omega^2$ by
$\Lambda=E^\varphi/\sqrt{|E^x|}$, $P_\Lambda= -\sqrt{|E^x|}K_\varphi$,
$R=\sqrt{|E^x|}$ and $P_R=-2\sqrt{|E^x|} K_x -E^\varphi
K_\varphi/\sqrt{|E^x|}$ where $P_\Lambda, P_R$ are the momenta
canonically conjugate to $\Lambda$ and $R$ respectively, $x$ is the
radial coordinate and $d\Omega^2=d\theta^2+\sin^2\theta
d\varphi^2$. We will take the Immirzi parameter and Newton's constant
to be unity.  

The total Hamiltonian (after a rescaling and an integration by parts) 
is given by
\begin{eqnarray}
H_T &=&\int dx \bigg[ -N'
  \bigg(-\sqrt{\vert
      E^x\vert}\left(1+K_\varphi^2\right)\nonumber\\
&&+\frac{\left[\left(E^x\right)'\right]^2\sqrt{\vert E^x\vert}}{4
          \left(E^\varphi\right)^2}
+ 
F(r)p\, \Theta(x-r)+2M\bigg)\nonumber\\
&& 
+ N^x \left[-
(E^x)' K_x +E^\varphi K_\varphi' - p\,
\delta\left(x-r\right)\right]\bigg]\nonumber\\
&&+N_-M + \frac{N_+}{2} \left[F(r)p+M\right], 
\end{eqnarray}
with $\Theta$ the Heaviside function, $r$ the position of the shell
and $p$ its canonical momentum.  Following Louko et al. \cite{louko}
we took the coordinates $x,r$ to have the $(-\infty,\ldots,\infty)$
range and assumed that the variables have the usual fall-offs in
asymptotic radial coordinates. We are including the usual boundary term
at spatial infinity with $N_+$ the lapse there $N_-$ the lapse at the
other end of the manifold extended beyond where the singularity in the
classical theory used to be and, as we will see, is removed by loop
quantum gravity like in the static case \cite{spherical}. We are
allowing the possibility of a pre-existing black hole of mass
$M$. Also, $ F(r) = \sqrt{|E^x|} \left(\eta
  \left(E^x\right)'\left(E^\varphi\right)^{-2} + 2 K_\varphi
  \left(E^\varphi\right)^{-1} \right)\vert_{x=r}.$ $N$ is a rescaled
lapse, $N=N_{\rm orig} E^\varphi/\left(E^x\right)'$ with $N_{\rm
  orig}$ the original unrescaled lapse and the shift is also changed
$N^x = N^x_{\rm orig} + 2 N_{\rm orig}
K_\varphi\sqrt{|E^x|}/\left(E^x\right)'$ with prime denoting derivative
with respect to the radial coordinate $x$. The parameter $\eta=\pm 1$
is the sign of the momentum, depending on it one will have shells that
are either outgoing or ingoing if one is outside the black hole.


Notice that $m\equiv F(r)p/2$ is a Dirac observable, which can be verified by
direct calculation of its Poisson brackets with the constraints. Also,
taking into account the falloff of the gravitational variables at
spatial infinity the constraint indicates that the ADM mass of the
space-time is $M+m$ (if there is no pre-existing black hole in the
space time and all the mass is provided by the shell we have
$M=0$). The lapse at infinity $N_+$ may be written as usual
\cite{kuchar} in terms of the proper time at infinity $\tau$ as
$N_+=\dot{\tau}$. It can also be verified that the Hamiltonian for
gravity coupled to a shell has an Abelian algebra with itself. One can
also check that the Poisson brackets with the diffeomorphism
constraint are the usual ones.

There also exists a Dirac observable, 
$
  V \equiv -\int_r^\infty dy \left[-2 F^{-1}(y) +\eta\left(1+2
      (M+m)/y\right)\right] +\tau -\eta\left[r+2 (M+m) \ln(r/(2(M+m)))\right],
$
and it can be checked that it has vanishing Poisson brackets with all
the constraints and is canonically conjugate to $m$. The observable
$V$ is associated with
the Eddington--Finkelstein coordinate $v$ of an observer at
scri
from which the shell is incoming or exiting.

For the quantization, we choose the same Hilbert space as in the test
shell case, which is a direct product of the Hilbert space of vacuum
spherically symmetric gravity and $L^2$ functions for the shell. As in
previous papers for the gravitational part we consider linear
combinations of products of cylindrical functions of the form,
\begin{eqnarray} T_{g,\vec{k},\vec{\mu}}(K_x,K_\varphi) &=&\prod_{e_j\in
g} \exp\left(\frac{i}{2} k_{j} \int_{e_j} dx\,K_x(x)\right)\nonumber\\
&&\times \prod_{x_j\in g} \exp\left(\frac{i}{2} \mu_{j} K_\varphi(x_j) \right),
\end{eqnarray} where the label $k_j\in\mathbb{Z}$ is the valence
associated with the edge $e_j$, and $\mu_j\in\mathbb{R}$ the valence
associated with the vertex $x_j$ (usually called ``coloring'').

We adopt a representation for the point holonomies as quasi-periodic
functions (for an alternative choice see \cite{sphericaljavi}), so
that the labels $\mu_j$ belong to a countable subset of the real line
with equally displaced points. We take $\mu_j=2\rho
(l_j+\delta_j)$ with $l_j$ an integer, $\delta_j$ between $0$ and
$1$ and $\rho$ is the polymerization parameter. The kinematical
Hilbert space associated with a given graph $g$ is, 
$ {\cal H}^g_{\rm kin}={\cal H}_{\rm kin}^M\otimes{\cal
H}_{\rm kin}^{\rm sh}\otimes \left[\bigotimes_{j=1}^n\ell^2_j\otimes
\ell_{\delta_j}^2\right],$
where $\ell_{j}^2$ and $\ell_{\delta_j}^2$ denote the space of square
summable functions of $k_j$ and $\mu_j$, respectively, ${\cal H}_{\rm
  kin}^M$ is the Hilbert space of square summable functions of the
ADM mass and ${\cal H}_{\rm kin}^{\rm sh}$ is the Hilbert space associated
with the shell variables, square integrable functions of $r$.

The full Hilbert space is equipped with the inner product
$ \langle
g,\vec{k},\vec{\mu},M,r |g',\vec{k}',\vec{\mu}',M',r'
\rangle=\delta(M-M')\delta(r,r')\delta_{\vec{k},\vec{k}'}\delta_{\vec{\mu},\vec{\mu}'}\delta_{g,g'},
$ where $\delta_{g,g'}$ is equal to the unit if $g = g'$
or zero otherwise, and similarly for $\vec{k}$ and $\vec{\mu}$.

On this space we have several well defined basic operators. The mass
and triads act multiplicatively and the operators associated with the
connection variables are holonomies in the case of $K_x$ and point
holonomies in the case of $K_\varphi$. Explicit expressions are given
in \cite{sphericaljavi}.

We write the Hamiltonian constraint as,
$
  \tilde{H}(x) = H_g(x)+ F(r)p\, \Theta(x-r).
$
The operator $\hat{E}^\varphi$ only acts at the vertices as the point
holonomies for $K_\varphi$ only have support there. This leads us to
consider a Hamiltonian constraint that only acts at the vertices, just
like in the full theory, 
\begin{eqnarray}
\hat{H}(x_j)  &:=&   \mathbf{H}^{\rm g}_j + \frac{1}{2} (\boldsymbol{\theta}_j \; \widehat{F(r)p}  + \widehat{F(r)p} \;  \boldsymbol{\theta}_j)\nonumber\\
&  =&   \mathbf{H}^{\rm g}_j + \frac{1}{2} \sum_i \mathbf{F}_i (\boldsymbol{\theta}_j \mathbf{X}_i + \mathbf{X}_i \boldsymbol{\theta}_j),
\end{eqnarray}
where $\boldsymbol{\theta}_j$ and $\mathbf{X}_j$ are
operators acting on the wavefunction of the shell and defined as,
\begin{eqnarray}
\boldsymbol{\theta}_j \psi(r) & :=&
  \int_0^{\epsilon_j} d \epsilon \theta(x_j+\epsilon-r)  \; \psi(r), \\
\boldsymbol{\delta}_j \psi(r)  &:=& 
\int_0^{\epsilon_j} d \epsilon \delta(x_j+\epsilon-r) \;  \psi(r), \\
\mathbf{X}_j  & :=& \frac{1}{2}( \boldsymbol{\delta}_j  \hat{p} +
\hat{p} \boldsymbol{\delta}_j), \quad
\hat{p} \psi (r)  := -i \partial_r \psi(r), 
\end{eqnarray}
and $\epsilon_j$ is the spacing of the vertices
$\epsilon_j=x_{j+1}-x_j$. The gravitational and shell parts 
of the Hamiltonian
are, 
\begin{align}
  \mathbf{H}^{\rm g}_j &= \widehat{H_{\rm g}(x_j)}=
\hat{b}_j \left(-1-\widehat{K_\varphi^2}(x_j)+{\hat{a}_j^2 \widehat{[1/E^\varphi]^{2}}(x_j)
        }+2\hat M\right),\\
\mathbf{F}_j&=  \widehat{F(x_j)}=2\, \hat{b}_j\left(\hat{a}_j
  \widehat{[1/E^\varphi]^{2}}(x_j) + 
\widehat{[K_\varphi/E^\varphi]}(x_j)\right).
\end{align}
They can all be written in terms of the elementary operators,
\begin{eqnarray}
  \widehat{K_\varphi^2}(x_j)&=&
\frac{\widehat{\sin\left(\rho K_\varphi(x_j)\right)}}{\rho}
\hat{E}^\varphi(x_j)
\frac{\widehat{\sin\left(\rho K_\varphi(x_j)\right)}}{\rho}
\hat{E}^\varphi(x_j)^{-1},\nonumber\\
\widehat{[K_\varphi/E^\varphi]}(x_j)&=& 
\frac{\widehat{\sin\left(\rho K_\varphi(x_j)\right)}}{\rho}
{\widehat{\cos\left(\rho K_\varphi(x_j)\right)}}
\hat{E}^\varphi(x_j)^{-1},\nonumber\\
\widehat{[1/E^\varphi]^{2}}(x_j) &=&
{\widehat{\cos\left(\rho K_\varphi(x_j)\right)}}
\hat{E}^\varphi(x_j)^{-1}
{\widehat{\cos\left(\rho K_\varphi(x_j)\right)}}
\hat{E}^\varphi(x_j)^{-1},\nonumber\\
\hat{a}_j&=&
\frac{\eta}{2}\left(\hat{E}^x(x_j)-\hat{E}^x(x_{j-1})\right),\quad
\hat{b}_n=\sqrt{|\hat{E}^x(x_j)|},\nonumber
\end{eqnarray}
where as usual we have polymerized the $K_\varphi$ variable but also
introduced a holonomy correction in $E^\varphi$ such that the change
is equivalent to a canonical transformation at the classical
level. There one has that $\{K_\varphi(x),E^\varphi(y)\}=\delta(x-y)$
and also that
$
  \left\{\frac{\sin\left(\rho K_\varphi(x)\right)}{\rho},
  {\cos\left(\rho K_\varphi(y)\right)}^{-1} E^\varphi(y)\right\}=\delta(x-y).
$
So we replace $K_\varphi$ by the sines and $E^\varphi$ by itself times
the inverse cosine in all expression, that is the canonical
transformation. This polymerization preserves the constraint algebra
and leads to the original classical theory when $\rho\to
0$. Interestingly it can also be extended to matter fields, like
scalar fields and it preserves the constraint algebra.  In a rather
lengthy but straightforward calculation it can be shown that for the
case of the shell the constraint has an Abelian commutator algebra
with itself also at the quantum level.

Unlike the static vacuum case, we do not at the moment know how to
find in closed form the space of physical states associated with the
constraints we defined. However, with the constraints represented by well defined
operators, we can recognize the quantum observables of the model
$\widehat{O(z(x))}$, $\widehat{M}$, $\widehat{\tau}$, 
$\widehat{F p}$ and $\widehat{V}$ ($z(x)$ is
a function of the radial variable that takes values in the interval
$[-1,1]$ that enters the definition of $\hat{E}^x$, we give more
details below, for additional details see \cite{sphericaljavi}).  The
former is present in the vacuum theory already and corresponds to the
diffeomorphism invariant content of $\hat{E}^x$, or in other words
that diffeomorphisms in one dimension cannot change the order of the
vertices of the spin network. The middle ones are the ADM mass and its
canonically conjugate momentum, the time at infinity. 
The two latter ones are associated with
the shell. They have the commutator, $ \left[\hat{m},\hat{V}\right]
= i\hbar.$ In terms of them we would like to express dynamical
variables of the problem as parameterized Dirac observables (evolving
constants of the motion) and study the physics in the quantum regime.
It should be emphasized that we do not have at the moment an explicit
self adjoint implementation of the observables associated with the
shell. We will assume one exists and, given their algebra, we can 
characterize the eigenstates of the complete set of observables and
define parameterized Dirac observables based on them.

Following similar steps to the ones we carried out in the vacuum case
\cite{hawking} leads to explicit expressions for the components of the
metric as functions of the Dirac observables and classical
(functional) parameters
$K_\varphi$ and $z(x)$. We are interested in studying the whole
space-time so we choose the horizon penetrating Eddington--Finkelstein
coordinates \cite{louko}. For that purpose we consider that $E^x$ is time-independent, what fixes $N^x=0$, and
a relationship between $K_\varphi$
and $E^x$ 
given by 
\begin{eqnarray}
  K_\varphi &=&
\frac{R_S\left[\Theta\left(x\right)-\Theta\left(-x\right)\right]}{\sqrt{|E^x|}\sqrt{1+\frac{R_S}{\sqrt{|E^x|}}}},
\end{eqnarray}
where 
\begin{eqnarray}
R_S &=& 2M + 2m\Big[\Theta\left(x\right) \Theta\left(\sqrt{|E^x|}+(t-V)\right) \nonumber\\
&&+ \Theta\left(-x\right) \Theta\left(\sqrt{|E^x|}-(t-V)\right)\Big].
\end{eqnarray}
We will concentrate from now on in the case
without a pre-existing black hole, i.e. $M=0$.  The expression for
$K_\varphi$ 
reduces to ordinary Eddington--Finkelstein
outside the shell in $x>0$ and satisfies
$\dot{K}_\varphi=\left\{K_\varphi,H_T\right\}$. Besides, one can check 
that $N=1/2$ and
\begin{eqnarray}
  E^\varphi &=&\frac{\left(E^x\right)'}{2}\sqrt{1+\frac{R_S}{\sqrt{|E^x|}}}.
\end{eqnarray}

These
expressions take values for $x<0$ that allow to extend the
parameterized observables to the region beyond the singularity. Since
generically we will be interested in considering superpositions of
states of different eigenvalues of $\hat{m}$ in order to approximate
semiclassical space-times, it is convenient to promote the above
expression to a quantum identity so it automatically adjusts to
each state.  This gauge choice extends the kind of gauge fixings up to
now considered where $K_\varphi$ was a c-number. We now consider a
choice that also depends on the observables. The conservation of the
gauge fixing ensures the consistency of the choice.  The resulting
expressions for the metric are (for reasons of space we give only one
component, the others are similar in nature),
\begin{eqnarray}
g_{tx}&=&-\frac{R_S}{2}\frac{\left(E^x\right)'}{|E^x|}\left[\Theta\left(x\right)-\Theta\left(-x\right)\right].
\end{eqnarray}
All the relevant quantities can be 
readily turned into well defined quantum operators. The operator
associated with $\hat{E}^x$ is
\begin{eqnarray}
{\hat{E}^x(x_j) }
|g,\vec{k},\vec{\mu},r\rangle &=& \widehat{O(z(x_j))}|g,\vec{k},\vec{\mu},r\rangle\nonumber\\&\equiv&
\ell_{\rm Pl}^2 k_{{\rm Int}(z(x_j)v)}
|g,\vec{k},\vec{\mu},r\rangle, 
\end{eqnarray}
where ${\rm Int}$ refers to the integer part and $z(x)$ is an
arbitrary function taking values in $[-1,1]$ whose choice determines
the radial coordinate chosen, $n=2v+1$ is the number of vertices of
the spin network considered (we work in a finite domain of length $L$
to avoid dealing with asymptotic issues involving spin nets). We
recall that $x_j$ is the radial coordinate of the $j$-th point in the
spin network. So the Dirac observables are specified in terms of two
parameters, one of them functional, $z(x)$ and $t$. Notice that the
resulting metric is time-dependent. To understand this time dependence
we recall that in Eddington--Finkelstein coordinates time is related
to the position of the shell by $t+x_r=V$ with $x_r$ the radius of
the shell, $E^x(r)=x_r^2$, and $V$ constant along the shell. The resulting picture of
the geometry is that of an ingoing shell that forms a black hole
traversing its event horizon, going towards $x=0$ and continuing
beyond to the region of $x<0$ as we observed in the case of test
shells \cite{testshell}. 

The above physical description of the metric is qualitative, based on looking at the
classical expressions of the metric as Dirac observables. Let us
discuss how this vision can be implemented in detail in the quantum
theory. 
We will consider a situation with 
states $\vert \vec{k}, m\rangle$ which are eigenstates of $\hat{O}$ and $\hat{m}$,
constituting a basis,
and form a wavepacket centered in $m_0, V_0$ of width $\sigma_m$.  We
proceed to construct a semiclassical solution based on a spin network
with $n$ vertices, $i=-v\ldots v$. Although there are infinitely many
possible choices for the position of the vertices we will choose a
simple one that leads to a good semiclassical behavior. We will place
the vertices of the spin network at radial positions $x_i=(i+1)\Delta$
and $\Delta$ a spacing bounded below by the quantization of areas
$\Delta> \ell_{\rm Planck}^2/(2x_r)$. We have for $i\ge0$ that $x_i\in
[\Delta, L]$ with $L=\Delta (v+1)$. Also,  $z(x)= x/L$ and
$k_i={\rm Int}(x_i^2/\ell_{\rm Planck}^2)$. In addition $
\left(E^x_i\right)' =\ell_{\rm Planck}^2 \frac{\left(
    k_i-k_{i-1}\right)}{\Delta} \sim \left(2i+1\right)\Delta.  $ On
the other hand, for $i<0$ we have that $k_i=-{\rm
  Int}(i^2\Delta^2/\ell_{\rm Planck}^2)$ and $\left(E^x_0\right)'
=\ell_{\rm Planck}^2 \left(k_0-k_{-1}\right)/\Delta = 2\Delta$ and it is
non-vanishing $i$'s where $k_i$, i.e. $E^x$, changes sign at $x=0$. 
The allowed spin networks need to exclude the value $k_0=0$ in order to 
ensure that the description is singularity free. The fact that both 
the Hamiltonian constraint and the observables are independent of 
$K_x$ (conjugate to $E^x$) ensures that this condition is a consistent 
restriction on the physical Hilbert space. Notice that $K_x$ is only present in 
the diffeomorphism constraint that does not change the sequence of 
values $\vec k$. With these
assumptions the result of the quantum construction is essentially a
discretization of the above classical expressions of the metric on a
lattice determined by a given spin network. They are extremely well
approximated by their classical counterparts in regions away from where
the classical singularity used to be. Close to it, such expressions
are sensitive to the lattice chosen. However, to have good
semiclassical behavior in the regions far away from the singularity,
it is best to consider superpositions of spin networks with different
values of $\vec{k}$ and $m$. This will imply that there are states 
providing a description of the region close
to where the singularity was where there will be large fluctuations of the
metric and therefore the singularity is replaced by a highly quantum but 
regular region that is not well approximated by a semiclassical space-time.

Because we are not including fields, and therefore Hawking radiation,
one ends up with a black hole that resembles the eternal black hole we
discussed in the past \cite{spherical}, with an interior region (which can 
be viewed as a ``baby universe'') beyond where the singularity was that is
disconnected from the exterior. In a more realistic approximation
involving Hawking radiation, the black hole will evaporate. Presumably
in that case the region where the shell emerges will coincide with the
exterior of the black hole. It is likely that we could approximate
large portions of such space-time with the results of this paper,
including the emergence of a shell from the highly quantum region that
has been conjectured in scenarios such as that of
\cite{asbo,rovelli,gapu,garay}. We cannot however provide a precise picture, in
particular, the time in which the shell emerges cannot be computed in
this approach, so generically for the outgoing shell one would have
$-x_r+t+t_0=V$ with $t_0$ a constant.

Summarizing, we have provided a detailed model for quantum collapse of
matter and formation of a black hole in a quantum geometry. It
provides insights into possible scenarios of black hole formation and
evaporation.

This work was supported in part by Grant No. NSF-PHY-1305000, ANII
FCE-1-2014-1-103974, funds of the Hearne Institute for Theoretical
Physics, CCT-LSU, and Pedeciba. J.O also acknowledges the projects
MICINN/MINECO FIS2011-30145-C03-02 and FIS2014-54800-C2-2-P (Spain).

\end{document}